\documentclass[aps,pre,twocolumn,amsmath,superscriptaddress,amssymb,showpacs]{revtex4}
\usepackage{graphicx}
\usepackage{bm}
\usepackage{amssymb}
\begin{document}
\title{Relaxation in statistical many-agent economy models} 
%

\author{Marco Patriarca}
\thanks{Corresponding author}
\email{marco.patriarca[at]gmail.com}
\affiliation{Institute of Theoretical Physics, Tartu University, T\"ahe 4,
51010 Tartu, Estonia}

\author{Anirban Chakraborti}
\email{achakraborti[at]yahoo.com}
\affiliation{Department of Physics, Banaras Hindu University, Varanasi-221005,
India}

\author{Els Heinsalu}
\email{ehe[at]ut.ee}
\affiliation{Institute of Theoretical Physics, Tartu University, T\"ahe 4,
51010 Tartu, Estonia}

\author{Guido Germano}
\email{germano[at]staff.uni-marburg.de}
\homepage{http://www.staff.uni-marburg.de/~germano}
\affiliation{Department of Chemistry, Philipps-University Marburg,
35032 Marburg, Germany}

\begin{abstract}
We review some statistical many-agent models of economic and social systems
inspired by microscopic molecular models and discuss their stochastic
interpretation.
We apply these models to wealth exchange in economics and study how the
relaxation process depends on the parameters of the system, in particular on
the saving propensities that define and diversify the agent profiles.
\end{abstract}



\maketitle

\section{Introduction}
One might question how theories that try to explain the physical world of
elementary particles, atoms, and molecules can be applied to understand the
social structure in its complexity and the economic behavior of human beings:
is it possible to describe the behavior of people with simple models? Is it even
possible to identify and quantify the nature of the interactions between them?
Even though it is still difficult to find answers to these questions, during the past decade physicists have made attempts to study problems related to economics, {\it the social science that seeks to analyze and describe the production, distribution, and consumption of wealth}~\cite{EB-economics}.

Here we will not try to review all these attempts, rather we briefly describe
what we name \emph{statistical many-agent models}. In these models, economic
activity is described as a flow of wealth
between basic units, referred to as agents, representing e.g.\ individuals or
companies. Each of the $N$ agents $\{1,\ 2,\ \dots,\ i,\ \dots,\ N\}$
has a wealth $x_i$, that changes in time as agents exchange wealth between
each other, according to the trading rules detailed in Sec.~\ref{review}.
These underlying trading rules only depend on one set of parameters, namely the
saving propensities $\{\lambda_i\}$, with $0 \le \lambda_i < 1$. Statistical
many-agent models describe closed economy systems and can reproduce some
features of wealth distributions, such as an exponential at intermediate
values of wealth and a power law at high values. In a particular model, all
agents have the same global $\lambda$; in a more general model, different
agents have different values of $\lambda$.

If the saving propensity is equal for all the units (global saving propensity
models), $\lambda_i \equiv \lambda$, the equilibrium wealth density $f(x)$ is
given by a $\Gamma$-distribution,
\begin{eqnarray}
\label{gamman}
f(x) &=& \beta\,\gamma_n(\beta x) = \frac{\beta}{\Gamma(n)} \, (\beta x)^{n-1}
\exp( - \beta x ) \, , \\
n &=& \frac{D}{2} = \frac{1 + 2\lambda}{1 - \lambda} \, .
\end{eqnarray}
Here $n$ is a real number in the interval $[1,\infty)$ and $D = 2n$ can be
considered as the effective dimension of the system: in fact the distribution
$\gamma_n(\beta x)$ is just the Maxwell-Boltzmann distribution for the kinetic
energy $x$ of a gas in thermal equilibrium at a temperature $\beta^{-1}$ in a
$D$-dimensional space. Thus the parameter $\beta$ can be interpreted as the
inverse temperature: consistently with the equipartition theorem, $\beta^{-1}
= 2 \langle x \rangle / D = \langle x \rangle / n$, where the constant $\langle
x \rangle$ is the average wealth $\langle x \rangle = X / N$ and $X =
\sum_{i=1}^N x_i$ is the total wealth. In an economic system, temperature is
proportional to the fluctuations of wealth around its average value. The model
with a single global saving propensity describes well wealth distributions at
intermediate values of $x$, but cannot predict power laws at large $x$.

If $\lambda$ is uniformly distributed among the agents according to a given
density $\phi(\lambda)$ (distributed saving propensity models), then one
finds, under quite general conditions on the shape of $\phi(\lambda)$, an
exponential law at intermediate values of $x$ and a robust Pareto law,
\begin{equation}
\label{pareto}
f(x) \propto x^{-1-\alpha},
\end{equation}
with $\alpha \ge 1$, at large $x$. This power law was suggested by Vilfredo
Pareto \cite{Pareto1897a} more than a century ago to describe the tail of
wealth distributions and is usually found to be characterized across various
countries by a Pareto exponent $\alpha \approx 3/2$.

The basic exchange laws underlying some of these models are reviewed in
Sec.~\ref{review}.
In Sec.~\ref{relaxation} we focus on the relaxation to equilibrium generated
by the exchange laws of models with uniformly distributed $\lambda$ and
illustrate its dependence on the saving propensity $\lambda$ through some
examples. We also consider the corresponding relaxation time distributions and
discuss the relation between the Pareto exponent and the relaxation behavior of
the system. Results are summarized in Sec.~\ref{conclusions}.

\section{Many-agent models of a closed pure exchange economy}
\label{review}
\begin{figure}
\begin{center}
\includegraphics[width=.48\textwidth]{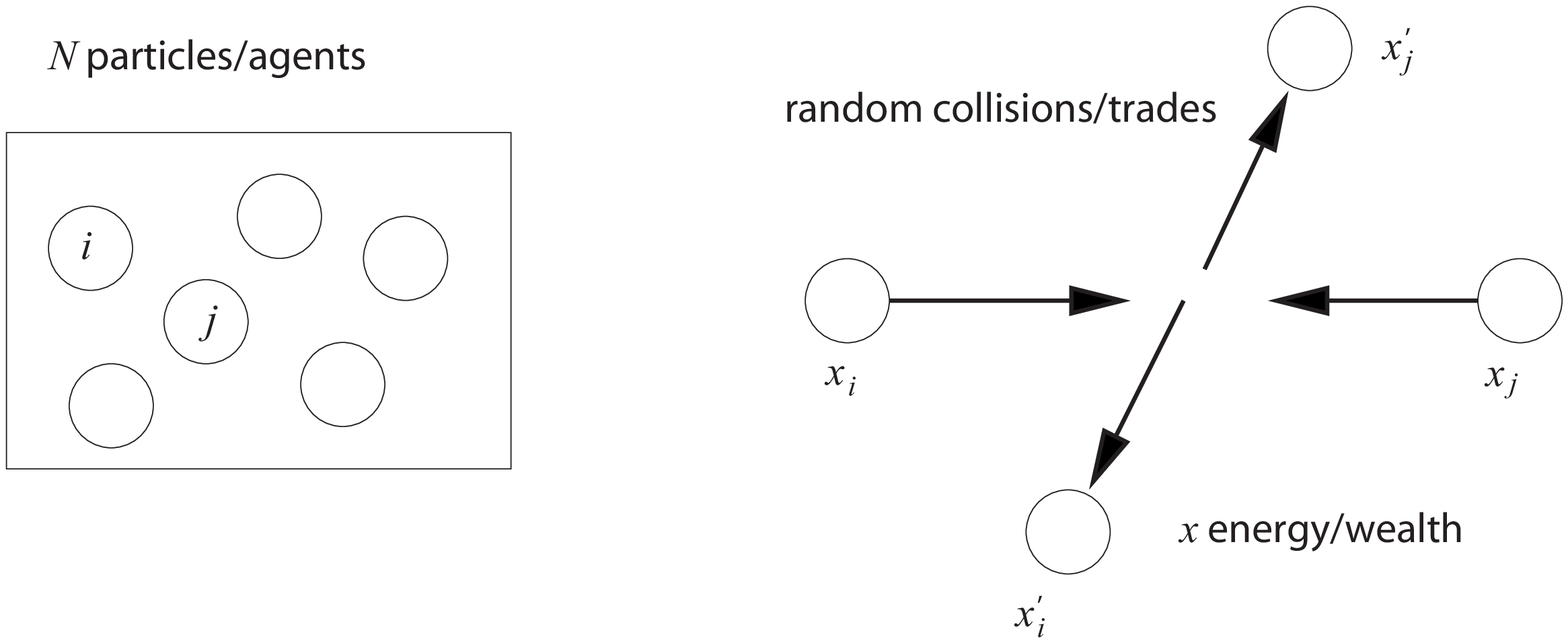}
\caption{Analogy between the minimal closed pure exchange economic model and a
classical isolated system of an ideal gas. In the latter case particles undergo
random elastic collisions and exchange a fraction of their kinetic energy,
while in the closed economy model agents perform random trades with each other
and exchange a fraction of their wealth according to some statistical rule.}
\label{fig:exchange}
\end{center}
\end{figure}

Our aim is to study a general many-agent statistical model of a closed economy
without growth (analogous to the kinetic theory model of ideal gases, see
Fig.~\ref{fig:exchange}), where $N$ agents exchange a quantity $x$, that we
have defined as wealth. The states of the agents are specified only by the
wealths $\{x_i\}$, while the total wealth $X=\sum_{i=1}^N x_i$ is conserved.
The evolution of the system is carried out according to the following algorithm:
at every time step two agents $i$ and $j$ are chosen randomly and an amount of
wealth $\Delta x$ is exchanged, so that the agent wealths $x_i'$ and $x_j'$
after the transaction are
\begin{eqnarray}
\label{basic0}
x_i' &=& x_i - \Delta x \, , \nonumber \\
x_j' &=& x_j + \Delta x \, .
\end{eqnarray}
Different transaction rules have been studied analytically or numerically by
various authors and are summarized here below.

\subsection{Basic model without saving: Boltzmann distribution}
\label{sec:basic}
A stochastic trading rule, that redistributes the wealths of two agents
randomly, was introduced in Ref.~\onlinecite{Dragulescu2000a},
\begin{eqnarray}
\label{basic1}
x_i' &=& \epsilon (x_i + x_j) \, , \nonumber \\
x_j' &=& \bar{\epsilon} (x_i + x_j) \, ,
\end{eqnarray}
where $\epsilon$ is a uniform random number in $(0,1)$ and $\epsilon +
\bar{\epsilon} = 1$. Eqs.~(\ref{basic1}) are equivalent to the trading
rule of Eq.~(\ref{basic0}) if $\Delta x = \bar{\epsilon} x_i - \epsilon x_j $.
In another version of the model, the money difference $\Delta x$ is assumed to
have a constant value independent of the two trading
agents~\cite{Bennati1988a,Bennati1988b,Bennati1993a}, i.e.\ $\Delta x =
\Delta x_0$. Both these forms for $\Delta x$ lead to a robust equilibrium
Boltzmann (or Gibbs) distribution,
\begin{equation}
\label{BD}
f(x) = \beta^{-1} \exp(-\beta x) \, ,
\end{equation}
with effective temperature $\beta^{-1} = \langle x \rangle = X/N$
\cite{Dragulescu2000a,Bennati1988a,Bennati1988b,Bennati1993a}.

\begin{figure*}
\begin{center}
\includegraphics[angle=0,width=.4\textwidth]{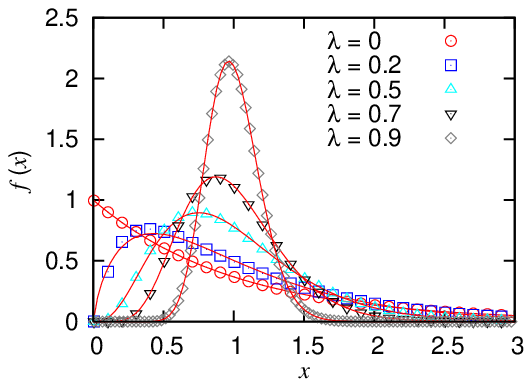}
\includegraphics[angle=0,width=.4\textwidth]{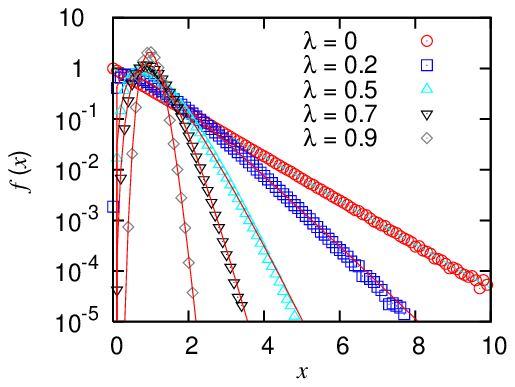}
\caption{Wealth probability density in linear (left) and semi-log (right) scale
for various global saving propensities $\lambda$. The exponential curve
$\lambda=0$ is the equilibrium solution for the basic model
(Sec.~\ref{sec:basic}), while the other curves correspond to a nonzero global
saving propensity (Sec.~\ref{global}).}
\label{fig:gamma}
\end{center}
\end{figure*}

\subsection{Model with a global saving propensity: Gamma distribution}
\label{global}
Introducing a saving criterion through a saving propensity parameter $0 \le
\lambda < 1$~\cite{Chakraborti2000a,Chakraborti2002a}
modifies the trading rule as follows:
\begin{eqnarray}
\label{sp1}
x_i' &=& \lambda x_i + \epsilon (1-\lambda) (x_i + x_j) \, , \nonumber \\
x_j' &=& \lambda x_j + \bar{\epsilon} (1-\lambda) (x_i + x_j) \, .
\end{eqnarray}
This model is similar to that introduced by John
Angle in 1983 on the basis of the Surplus Theory~\cite{Angle1983a,Angle1986a,Angle2006a},
which however differs both in the mathematical definition of the exchange rule and interpretation~\cite{Angle2006a}.
A closer comparison will be studied elsewhere, while here, for the sake of simplicity, we will focus on the trading rule (\ref{sp1}) and some of its generalizations for studying the relaxation
process.
The rule of Eq.~(\ref{sp1}) corresponds to the process defined by
Eq.~(\ref{basic0}) if
\begin{equation}
\Delta x = (1 - \lambda) ( \bar{\epsilon} x_i - \epsilon x_j ) \, .
\end{equation}
The parameter $\lambda$ represents the fraction of wealth saved before the
reshuffling takes place. The resulting equilibrium distribution is
qualitatively different from a simple exponential function, being a $\Gamma$-distribution~\cite{Patriarca2004a,Patriarca2004b}, see Eq.~(\ref{gamman}), that
has a mode $x_m>0$ and a zero limit for $x \to 0$, see Fig.~\ref{fig:gamma}.

\subsection{Models with a continuous distribution of saving propensities}
\label{continuous}
More realistic and interesting models are obtained when agents $i = 1,\ \dots,\
N$ are diversified by assigning them different saving propensities
$\lambda_i$~\cite{Angle2002a,Chatterjee2003a,Das2003a,Chatterjee2004a,Repetowicz2005a,Chatterjee2005a,Das2005a,Patriarca2005a}, e.g.\ with the $\lambda_i$
distributed uniformly on the interval $[0,1)$. The trading rule is then
\begin{eqnarray}
\label{sp2}
x_i' &=& \lambda_i x_i + \epsilon [ (1-\lambda_i) x_i + (1-\lambda_j) x_j ]
\, , \nonumber \\
x_j' &=& \lambda_j x_j + \bar{\epsilon} [(1-\lambda_i) x_i + (1-\lambda_j) x_j ]
\, ,
\end{eqnarray}
or, equivalently, can be formulated through Eq.~(\ref{basic0}) with
$\Delta x$ given by
\begin{equation}
\label{dx_sp2}
\Delta x = \bar{\epsilon} (1-\lambda_i) x_i - \epsilon (1-\lambda_j) x_j \, .
\end{equation}
Numerical simulations and theoretical considerations suggest that these models
relax toward a robust power law $\propto 1/x^{1 + \alpha}$ with a Pareto
exponent
$\alpha =1$ in the case of uniformly distributed $\lambda$ and with $\alpha >
1$ if the $\lambda$-density $\phi(\lambda) \to 0$ as $\phi(\lambda) \sim
(1-\lambda)^{\alpha-1}$ for $\lambda \to 1$. In the following we study the
relaxation process of models with uniformly distributed $\lambda$.

\section{Relaxation process}
\label{relaxation}
If a real economic system is characterized by a wealth distribution with a
certain shape, it is of great interest to know on which time scale the system
relaxes toward this distribution from a given arbitrary initial distribution of
wealth, and how the relaxation process depends on the system parameters, in
particular on the system size and the distribution of saving propensities.

In the simulations presented below, all agents start from the same initial
wealth $x_i(t = 0) = x_0 = 1$. The value $x_0$, due to the conservation of the
total wealth $X = \sum_{i=1}^N x_i$, also represents the global average value
of $x$ at any time $t$, i.e.\ $\langle x(t) \rangle \equiv \int x f(x) \, dx =
x_0 = X/N$. This setup is used to
model a more general situation where the initial conditions of the
agents are far from equilibrium.

\subsection{Relaxation to equilibrium as a function of system size}
\begin{figure}[ht]
\begin{center}
  \includegraphics[angle=0,width=.4\textwidth]{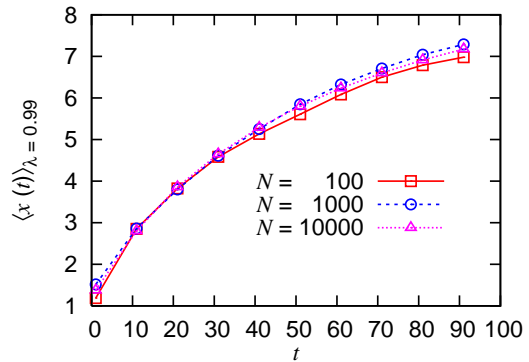}
  \caption{Average wealth $\langle x(t) \rangle_{\lambda = 0.99}$ versus the
rescaled time $t = T/N$ for systems with different number of agents $N = 100$,
$1000$, and $10\,000$, but the same saving propensity density $\phi(\lambda)$,
that uniformly partitions agents into $100$ subsets with $\lambda = 0.00,\
0.01,\ \dots,\ 0.99$. $T$ is the total number of trades.}
\label{fig:Compare-N}
\end{center}
\end{figure}

Before analyzing the dependence of the time scale on the saving propensity
distribution, we shortly consider its dependence on the number of agents $N$.
If time is measured by the number of transactions $T$, we find that the
time scale is proportional to the number of agents $N$: a system A that is $m$
times larger than a system B ($N_\mathrm{A} = m N_\mathrm{B}$) relaxes $m$
times slower than B. This is shown in Fig.~\ref{fig:Compare-N}, where the
average wealth $\langle x(t) \rangle_\lambda$ of the agent subset with
$\lambda=0.99$ is plotted for various systems with different values of $N$
versus the rescaled time $t = T/N$. However, the $\lambda$-density
$\phi(\lambda)$ is the same for all systems and uniformly partitions each
system into $100$ subsets with values $\lambda = 0.00,\ 0.01,\ \dots,\ 0.99$.

Here and in the following we define time $t$ as the ratio $t = T/N$ between the
total number of trades $T$ and the total number of agents $N$, i.e.\ what is
usually called a Monte Carlo cycle or sweep in molecular
simulation~\cite{Allen1989b}: in a Monte Carlo cycle, each agent
performs on average the same number of trades (actually two), in the same fashion as in
molecular dynamics each particle is moved once at every time step.
The results will not change if one of the two
agents involved in an exchange is selected sequentially, e.g.\ in the order of
its index $i = 1,\ \dots,\ N$, as is common practice in molecular simulations.
This ensures that every agent performs at least one trade per cycle and reduces
the amount of random numbers to be drawn.
The previous considerations suggest the introduction of a time unit $\tau_0$,
such that during any time interval $(t, t +\tau_0)$ \emph{all} agents perform
on average one trade (or the same number of trades). In this way the dynamics
and the relaxation process become independent of $N$. The existence of a
natural time scale independent of the system size provides a foundation for
using simulations of systems with finite $N$ in order to infer properties of
systems with continuous saving propensity distributions and $N \to \infty$.

\subsection{Relaxation to equilibrium as a function of saving propensity}
\label{relax2}
Relaxation in systems with constant $\lambda$ has already been studied in
Ref.~\onlinecite{Chakraborti2000a}, where a systematic increase of the
relaxation time with $\lambda$, and eventually a divergence for $\lambda \to
1$, was found: for $\lambda = 1$, no exchanges can occur, so that the system is
frozen. Here we consider systems with uniformly distributed $\lambda$.
In this case a similar behavior of the relaxation times is observed, broken
down to subsystems with similar values of $\lambda$. As discussed in detail in
Refs.~\onlinecite{Patriarca2005a,Bhattacharya2005a,Patriarca2006c}, the partial
wealth distributions of agents with a given value of $\lambda$ relax toward
different states with characteristic shapes $f_\lambda(x)$. The generic function
$f_\lambda(x)$ has a maximum and an exponential tail, thus closely recalling
the shape of a $\Gamma$-distribution. The corresponding average value is given
by $\langle x \rangle_\lambda \equiv \int x f_\lambda(x) \, dx =
k/(1-\lambda)$, where $k$ is a suitable constant determined through the
condition $\int \langle x \rangle_\lambda \, \phi(\lambda) \, d\lambda = X/N$;
$X$ is the total wealth of the system. Even if the partial distributions
decay exponentially with $x$, the sum of all partial distributions
results in a Pareto law at large values of $x$, i.e.\ $f(x) = \sum_\lambda
f_\lambda(x) \sim 1 / x^{1 + \alpha}$.
Numerical simulations clearly show that agents with different values of
$\lambda$ are associated to different relaxation times $\tau_\lambda$.

\begin{figure}[ht]
\begin{center}
\includegraphics[angle=0,width=.4\textwidth]{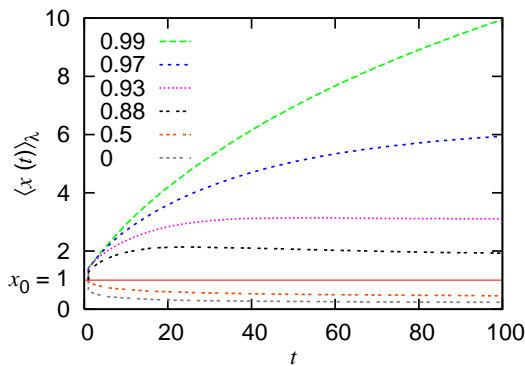}
\caption{ Mean wealth $\langle x(t) \rangle_\lambda$ versus time for various
$\lambda$. The $10^4$ agents are uniformly partitioned into $100$ subsets with
$\lambda = 0.01,\ 0.02,\ \dots,\ 0.99$. Higher $\lambda$ correspond to longer
relaxation times. The continuous line $\langle x \rangle = x_0$ partitions
agents into \emph{poor} ($x < x_0$) and \emph{rich} ($x > x_0$) ones.}
\label{fig:x-vs-t}
\end{center}
\end{figure}

Results are illustrated in Fig.~\ref{fig:x-vs-t} for a system of $N = 10^4$
agents uniformly partitioned into $100$ subsets with $\lambda = 0.01,\ 0.02,\
\dots,\ 0.99$: mean wealths of subsets corresponding to a value of $\lambda$
closer to 1 relax slower toward their asymptotic average wealth $\langle x
\rangle_\lambda \propto 1/(1-\lambda)$.

The average wealth $x_0$ allows to introduce a threshold that partitions
the system into \emph{poor agents}, with an asymptotic average wealth
$\langle x(t \to \infty) \rangle_\lambda < x_0$, and \emph{rich agents} with
$\langle x(t \to \infty) \rangle_\lambda > x_0$. The poor-rich threshold
$\langle x \rangle = x_0 = 1$ is represented as a continuous line in
Fig.~\ref{fig:x-vs-t} and corresponds to $\lambda \approx 0.75$ for
this particular example.

\begin{figure*}
\begin{center}
\includegraphics[angle=0,width=.4\textwidth]{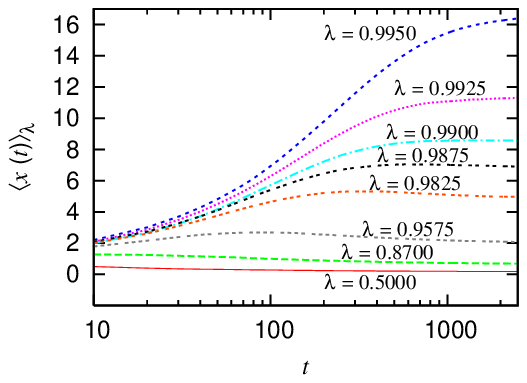}
\includegraphics[angle=0,width=.53\textwidth]{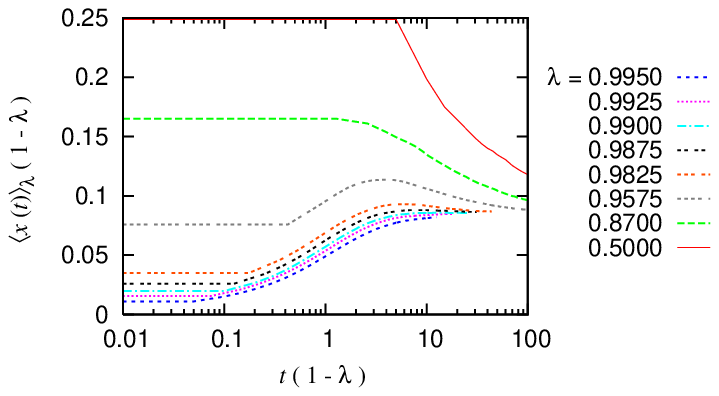}
\caption{
Left: average wealth $\langle x(t) \rangle_\lambda$ versus (log of) time for
the $\lambda$'s listed in the figure. The $10^4$ agents are partitioned into
$200$ subsets with $\lambda = 0.5000,\ 0.5025,\ \dots,\ 0.9975$. Right: same as
in the left figure after rescaling wealth and time by the factor $(1-\lambda)$,
inversely proportional to the mean wealth and proportional to the average
wealth exchange rate.}
\label{fig:compare1}
\end{center}
\end{figure*}

The differences in the relaxation process can be related to the different
relative wealth exchange rates, that by direct inspection of Eqs.~(\ref{sp2})
and (\ref{dx_sp2}) appear to be proportional to $1 - \lambda$. Thus, in
general, higher saving propensities are expected to be associated to slower
relaxation processes. A more detailed analysis can be carried out as shown in
Fig.~\ref{fig:compare1}: after the rescaling of time and wealth by the factor
$(1-\lambda)$, mean wealths corresponding to agents with different values of
$\lambda$ (Fig.~\ref{fig:compare1} left) appear to relax approximately \emph{on
the same time scale} and toward the same asymptotic value
(Fig.~\ref{fig:compare1} right).
In fact, the factor $(1-\lambda)$ is proportional to the wealth exchange rates
and, at the same time, through the condition of stationarity, determines the
equilibrium average wealth values $\langle x \rangle_\lambda =
k/(1-\lambda)$~\cite{Patriarca2005a}. Agents start from the same initial
condition $x_i(t = 0) = x_0 = 1$. In this case, in order to study in greater
detail the high saving propensity parameter region, that corresponds to
the high relaxation time region, the system of $N = 10^4$ agents has been
uniformly partitioned into $200$ subsets with saving propensities $\lambda =
0.5000,\ 0.5025,\ \dots,\ 0.9975$. Actually, this is not a uniform distribution
of $\lambda$ on $[0,1)$, since $\phi(\lambda) = 0$ for $\lambda < 0.5$, however
it does not matter because what counts is the high saving propensity parameter
interval.

\subsection{Relaxation time distribution}
The model with distributed saving propensities is completely specified by the
trading rules of Eqs.~(\ref{sp2}) and the set of saving propensities $\{\lambda_i\}$ of the
$N$ agents. In the case of a continuously distributed $\lambda$, a continuous
saving propensity density $\phi(\lambda)$ can be used in place of the discrete
$\lambda$-set, normalized so that $\int_0^1 \phi(\lambda) \, d\lambda = 1$.

Here we suggest a method to obtain the wealth as well as the relaxation
distribution directly from the saving propensity density $\phi(\lambda)$.
It follows from probability conservation that $\tilde{f}(\bar{x})d\bar{x} =
\phi(\lambda) d\lambda$, where $\bar{x}$ is a short notation for $\langle
x \rangle_\lambda$ and $\tilde{f}(\bar{x})$ is the density of the average
wealth values. In the case of uniformly distributed saving propensities,
one obtains
\begin{equation}
\label{f-phi}
\tilde{f}(\bar{x}) = \phi(\lambda) \frac{d\lambda(\bar{x})}{d\bar{x}}
= \phi\left(1-\frac{k}{\bar{x}}\right) \frac{k}{\bar{x}^2} \, .
\end{equation}
This shows that a uniform saving propensity distribution leads to a power law
$\tilde{f}(\bar{x}) \sim 1/\bar{x}^2$ in the (average) wealth distribution. In
general a $\lambda$-density going to zero for $\lambda \to 1$ as $\phi(\lambda)
\propto (1 - \lambda)^{\alpha-1}$ (with $\alpha \ge 1$) leads to the Pareto
law $\tilde{f}(\bar{x}) \sim 1/\bar{x}^{1 + \alpha}$ with Pareto exponent
$\alpha \ge 1$ as found in real distributions.

In a very similar way it is possible to obtain straightforwardly the associated
distribution of relaxation times $\psi(\tau)$ for the global relaxation process
through the relation between the relaxation time $\tau_\lambda$ and the agent
saving propensity: given that the time scale follows a relation $\tau_\lambda
\propto 1 / (1 - \lambda)$, then
\begin{equation}
\label{psi-phi}
\psi(\tau) = \phi(\lambda) \frac{d\lambda(\tau)}{d\tau} \propto
\phi\left(1 - \frac{\tau'}{\tau}\right) \frac{\tau'}{\tau^2} \, ,
\end{equation}
where $\tau'$ is a proportionality factor. Comparison with Eq.~(\ref{f-phi})
shows that $\psi(\tau)$ and $\tilde{f}(\bar{x})$ are characterized by power law
tails in $\tau$ and $\bar{x}$ respectively \emph{with the same Pareto
exponent}.

It is to be noticed, as discussed in Ref.~\onlinecite{Patriarca2005a}, that in
the parameter region $\lambda \to 1$, from which the main contributions to the
Pareto power law tail come, the widths of the generic equilibrium partial
distributions $f_\lambda(x)$ increase more slowly than the difference between
the mean values $\langle x \rangle_{\lambda'} - \langle x \rangle_\lambda$
corresponding to two agents with consecutive values of the saving propensity
$\lambda'$ and $\lambda$. This implies that at equilibrium and in the tail
of the distribution it is possible to resolve the mixture $\sum_\lambda
f_\lambda(x)$ into its components $f_\lambda(x)$ and to approximate the
current value of wealth $x(t)$ of a certain agent with saving propensity
$\lambda$ (that is actually a stochastic process) with the corresponding
average value, $\langle x \rangle_\lambda \approx x$, so that $\tilde{f}(x)
\approx f(x)$.

Finally we notice that an ensemble with a power law distribution of relaxation
times undergoes a slow relaxation process if the exponent of the relaxation
time distribution is smaller than two, so that a Pareto exponent larger
than two, as automatically generated by the model, seems to ensure a normal
relaxation.

\section{Conclusions}
\label{conclusions}
The relaxation process of statistical many-agent models of a closed pure
exchange economy, where trading is described as a flux of wealth between
different agents, has been found to be slower for agents with a larger saving
propensity parameter $\lambda$, who are also the agents resulting richer
at equilibrium. For a uniform $\lambda$-distribution, the relaxation time is
$\tau_\lambda \propto 1/(1-\lambda)$. Furthermore, a smooth distribution of
saving propensities leads to distributions of wealth and of relaxation times
characterized by power law tails with the same Pareto exponent $\alpha \ge 1$,
which ensures a fast relaxation toward equilibrium.

We also remark that if time is measured in Monte Carlo cycles, i.e.\ the
ratio between the total number of trades and the total number of agents, so
that \emph{every} agent performs on average two trades during a
cycle, the time evolution and the relaxation process are independent of the
system size, thus providing information on arbitrarily large systems.

\end{document}